\input{aipcheck}

\documentclass[
    ,final            
  ]
  {aipproc}

\layoutstyle{6x9}

\begin{document}

\title{Recent Developments in Neutron Star Thermal Evolution Theories and Observation(*)}
\footnote{(*) Invited paper presented in International Symposium
on Origin of Matter and Evolution of Galaxies (OMEG05), November
8-11, 2005, University of Tokyo, Tokyo, Japan}



\author{Sachiko Tsuruta}{
  address={Physics Department, Montana State University, Bozeman,
Montana 59717 USA} }

\begin{abstract}
Recent years have seen some significant progress in theoretical
studies of physics of dense matter. Combined with the
observational data now available from the successful launch of
{\it Chandra} and {\it XMM/Newton} X-ray space missions as well as
various lower-energy band observations, these developments now
offer the hope for distinguishing various competing neutron star
thermal evolution models. For instance, the latest theoretical and
observational developments may already exclude both nucleon and
kaon direct Urca cooling.  In this way we can now have a realistic
hope for determining various important properties, such as the
composition, superfluidity, the equation of state and stellar
radius. These developments should help us obtain deeper insight
into the properties of dense matter.
\end{abstract}

\maketitle


\section{Introduction}

The launch of the {\it Einstein} Observatory gave the first hope
for detecting thermal radiation directly from the surface of
neutron stars (NSs).  However, the temperatures obtained by the
{\it Einstein} were only the upper limits[1]. {\it ROSAT} offered
the first confirmed detections (not just upper limits) for such
surface thermal radiation from at least three cooling neutron
stars, PSR 0656+14, PSR 0630+18 (Geminga) and PSR 1055-52[2].
Recently the prospect for measuring the surface temperature of
isolated NSs, as well as obtaining better upper limits, has
increased significantly, thanks to the superior X-ray data from
{\it Chandra} and {\it XMM/Newton}, as well as the data in the
lower energy bands from optical-UV telescopes such as {\it Hubble
Space Telescope}. Consequently, the number of possible surface
temperature detections has already increased to at least seven[3].
Very recently {\it Chandra} offered an important upper limit to
PSR J0205+6449 in 3C58[4]. At the same time, more careful and
detailed theoretical investigation of various input microphysics
has been in progress[5]-[9]. The current paper is meant as a
progress report on these recent developments. Specifically, we try
to demonstrate that distinguishing among various competing NS
cooling theories has started to become possible, by careful
comparison of improved theories with new observations[3][10][11].

\section{Neutron Star Cooling Theories}

The first detailed cooling calculations[12] showed that isolated
NSs can be warm enough to be observable as X-ray sources for about
a million years. After a supernova explosion a newly formed NS
first cools via various neutrino emission mechanisms before the
surface photon radiation takes over. Among the important factors
which seriously affect the nature of NS cooling are: neutrino
emission processes, superfluidity of constituent particles,
composition, mass, and the equation of state (EOS)[3]. In this
paper, for convenience, the conventional, slower neutrino cooling
mechanisms, such as the modified Urca, plasmon neutrino and
bremsstrahlung processes, will be called `standard cooling'.  On
the other hand, the more `exotic' extremely fast cooling
processes, such as the direct Urca processes involving nucleons,
hyperons, pions, kaons, and quarks, will be called `nonstandard'
processes[3].

The composition of NS interior is predominantly neutrons with only
a small fraction of protons, electrons and muons when the interior
density is not high (the central density $\rho^c < \sim$ 10$^{15}$
gm/cm$^3$). For higher densities more `exotic' particles, such as
hyperons, pions, kaons and quarks, may dominate the central core.
Therefore, when the star is less massive and hence less dense, we
have a neutron star with the interior consisting predominantly of
neutrons (no `exotic' particles) and it will cool with the slower,
`standard' neutrino processes. On the other hand when $\rho^c$
exceeds the transition density to the exotic matter $\rho_{tr}$,
the transition from nucleons to `exotic' particles takes place.
Therefore, more massive stars, whose $\rho^c$ exceeds $\rho_{tr}$,
possess a central core consisting of the exotic particles. In that
case, the nonstandard fast cooling takes over. Note, however, that
if the proton fraction in the neutron matter is exceptionally
high, i.e., $> \sim 15\%$, very fast cooling can take place in a
NS without any exotic particles, through the nucleon direct Urca
process. This can happen for a certain type of EOS models which
allow such high proton concentration above a certain critical
density[13]. In order to include this option, in the following
discussion we will call any fast nonstandard process, an `exotic
process', rather than `a process involving exotic particles'. The
observational data suggest that there are at least two classes of
NSs, the hotter and cooler. The most natural explanation is that
the hotter stars are less massive and cool by slower cooling
processes, while the cooler ones are more massive and cool by one
of the fast nonstandard processes[3][10].

As the central collapsed star cools after a supernova explosion
and the interior temperature falls below the superfluid critical
temperature, T$^{cr}$, some constituent particles become
superfluid. That causes suppression of both specific heat (and
hence the internal energy) and all neutrino processes involving
the superfluid particles.  The net effect is that in the case of
fast nonstandard cooling, the star cools more slowly and hence the
surface temperature and luminosity will be higher at a given age
during the neutrino cooling era, due to the suppression of
neutrino emissivity. Therefore, nonstandard fast cooling will be
no longer so fast if the superfluid energy gap, which is
proportional to T$^{cr}$, is significant. In fact, if the gap is
large enough, nonstandard cooling is fully suppressed and the
cooling curve becomes essentially the same as the standard cooling
curve. Therefore, depending on the size of the energy gap, a
nonstandard cooling curve can lie anywhere between the standard
curve and the nonstandard one without superfluid suppression.

In addition to various neutrino cooling mechanisms conventionally
adopted in earlier calculations, recently the `Cooper pair
neutrino emission'[14][15] was `rediscovered' to be also important
under certain circumstances. This process takes place when the
participating particles become superfluid, and the net effect is
to enhance, for some superfluid models, the neutrino emission
right after the superfluidity sets in[15][16].


\begin{figure}
\begin{minipage}[t]{0.5\textwidth}
\includegraphics[height=.45\textheight, angle=270]{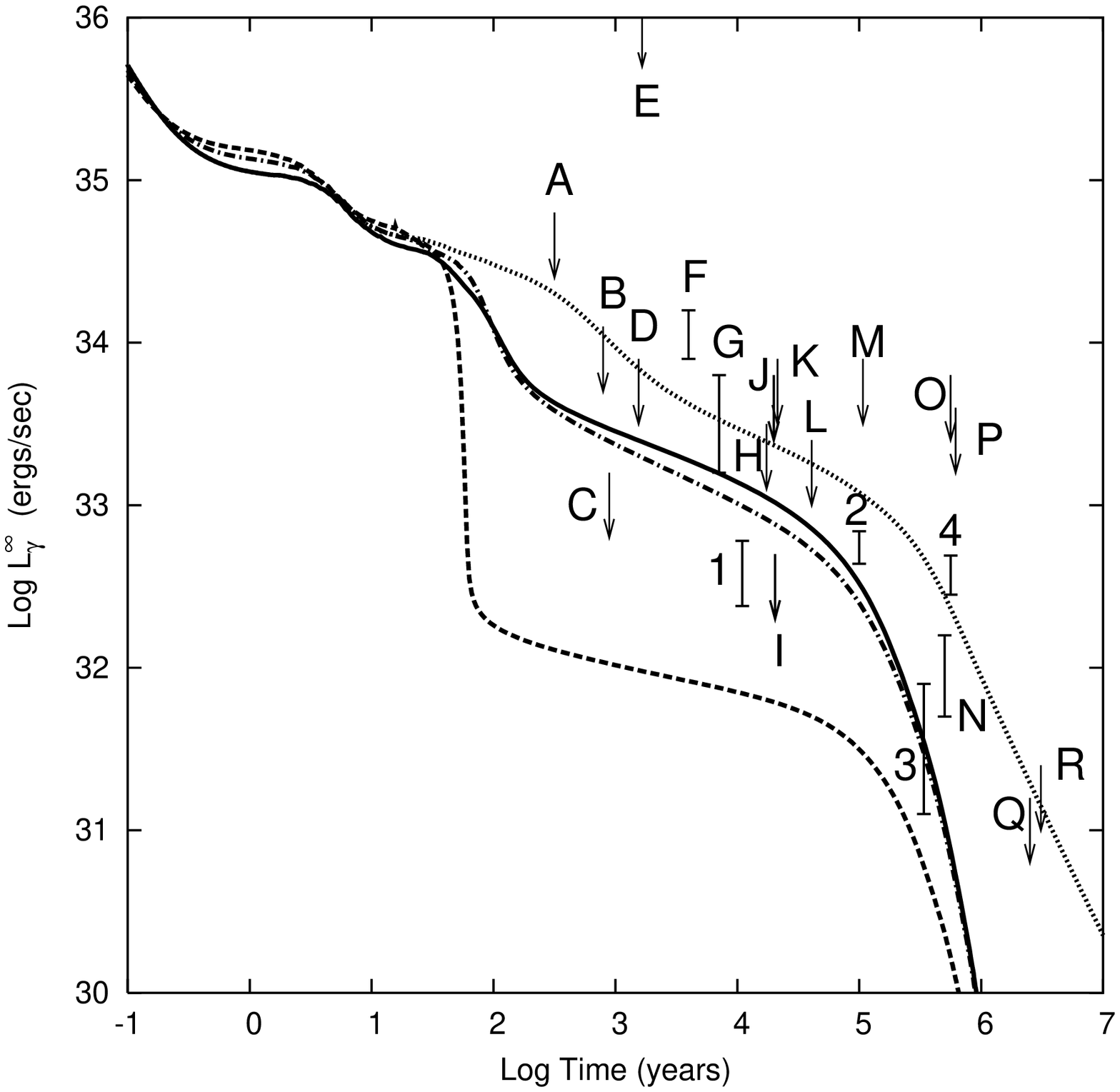}\end{minipage}
\begin{minipage}[t]{0.5\textwidth}
\includegraphics[height=.45\textheight, angle=270]{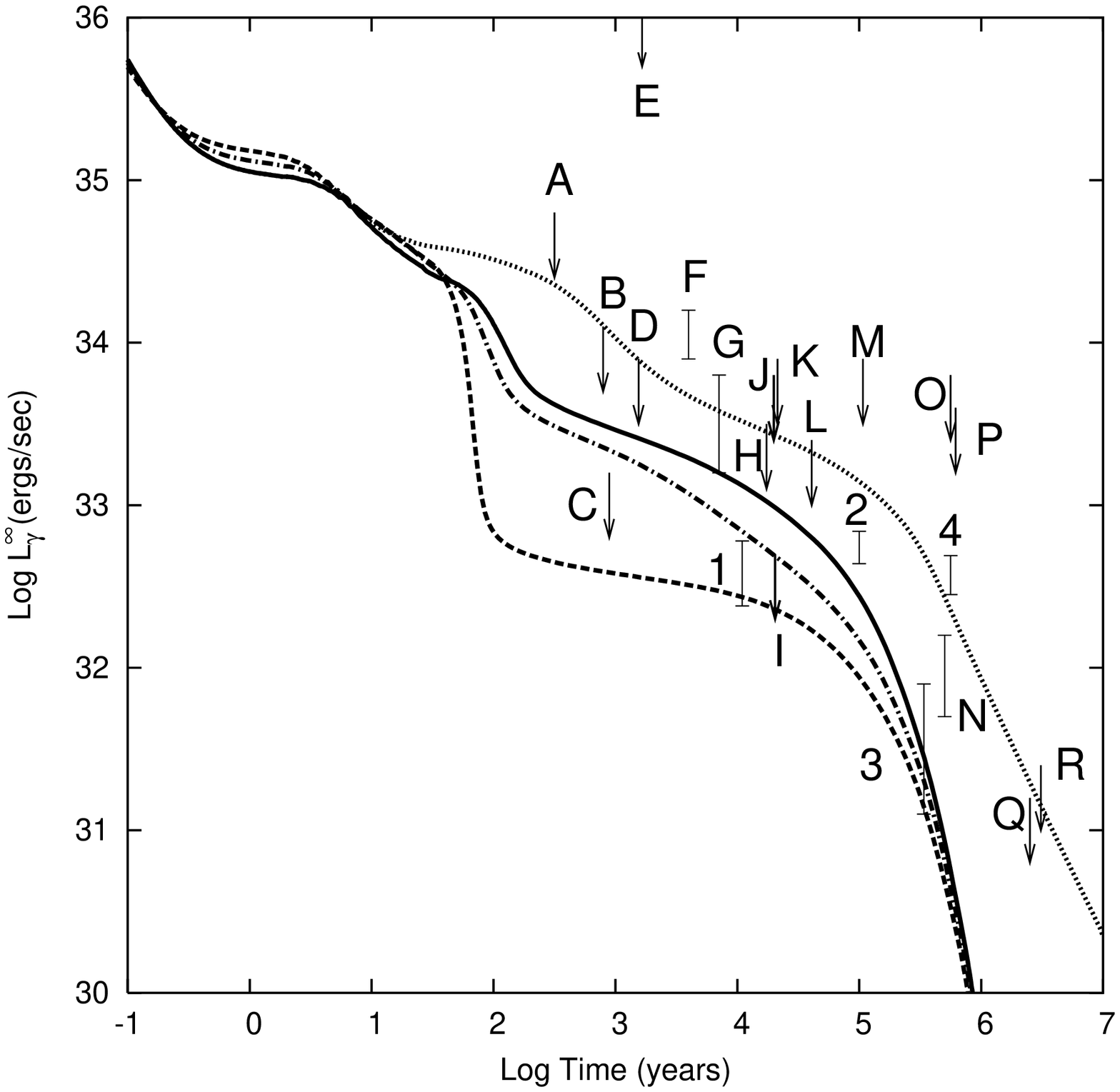}
\end{minipage}
\caption{Thermal evolution curves with the newest observational
data. In Fig. 1a (left panel) the dotted and solid curves refer to
the standard cooling of M = 1.4M$_\odot$ neutron stars with and
without heating, respectively, while the dot-dashed and dashed
curves are for hyperon cooling of 1.6 and 1.8M$_\odot$ stars,
respectively. In Fig. 1b (the right panel) the solid, dot-dashed
and dashed curves refer to pion cooling of 1.4, 1.6 and
1.8M$_\odot$ stars, respectively. In the same figure the dotted
curve refers to thermal evolution of a 1.4M$_\odot$ pion star with
heating. The vertical bars refer to temperature detection data
with error bars, while the downward arrows refer to the upper
limits. The more accurate detection data are sown with numbers,
for (1) the Vela pulsar, (2) PSR 0656+14, (3) Geminga, and (4) PSR
1055-52. The rest of the data shown are more rough estimates. Some
of more interesting among these are shown with letters, as (A) Cas
A point source, (B) the Crab pulsar, (C) PSR J0205+6449 in 3C58,
(F) RX J0822-4300, (G) 1E1207.4-5209, (I) PSR 1046-58, (N) RX
J1856-3754, and (R) PSR 1929+10. The complete list of all these
data sources is found in T06a,b[10][11]. Sources RX J0002+62 and
RX J0720.4-3125 are not shown because currently there are still
too large uncertainties including the age estimate. See the text
for further details.}
\end{figure}

\section{Most Recent Thermal Evolution Models}

Most recently we calculated NS thermal evolution\footnote{We adopt
the expression `thermal evolution' when we include not only
cooling but also heating.} adopting the most up-to-date
microphysical input and a fully general relativistic, `exact'
evolutionary code (i.e., without making isothermal
approximations). This code was originally constructed by Nomoto
and Tsuruta (1987)[17] which has been continuously up-dated. Our
input neutrino emissivity consists of all possible mechanisms,
including Cooper pair emission.  See Tsuruta et al. 2006a,b,
hereafter referred to as T06a,b[10][11] for the details. In the
models presented here, we consider thermal evolution of neutron
stars which possess a central core consisting of hyperons and pion
condensates at high densities, which we conveniently call hyperon
and pion stars. The results are summarized in Figure 1, where
surface photon luminosity which corresponds to surface temperature
(both to be observed at infinity), is shown as a function of age.

Fig. 1a (the left panel) shows thermal evolution of neutron (lower
mass) and hyperon (higher mass) stars. The critical transition
density from neutron to hyperon matter, $\rho_{tr}^Y$, is set at
4$\rho_0$, which is estimated by nuclear theories[7]. (Here
$\rho_0$ = 2.8 x 10$^{14}$ gm/cm$^3$ is the nuclear density.) For
$\rho < \rho_{tr}^Y$ we adopt the TNI6 EOS recently constructed
for neutron matter by Takatsuka et al 2006[7], while for $\rho
> \rho_{tr}^Y$ it becomes TNI6U, the same EOS but for hyperon
matter[7]. This EOS is medium in stiffness\footnote{Often an EOS
is referred to as being `stiff' when the consequent stellar model
is more extended and less dense, while it is referred to as being
`soft' if it is more compact and denser for a given mass.} and it
is very similar to the FP model adopted earlier, e.g. in Umeda et
al. 1994[18] and Umeda, Tsuruta and Nomoto 1995[19]. As the
superfluid model for hyperons we adopt the Ehime Model[7] and as
the neutron superfluid model the OPEG-B Model[6], both recently
constructed. The Cooper pair neutrino emissivity derived by
Yakovlev, Levenfish, and Shibanov (1999)[14] is adopted for both
neutrons and protons. For our heating calculations we adopt the
vortex creep heating model with the heating parameter K =
10$^{37}$ ergs m$^{-3/2}$ s$^2$, which is maximum in strength
according to theoretical estimates[19][20], and magnetic field B =
10$^{12}$ Gauss, reasonable for ordinary pulsars[21]. The other
input parameters are the same as in Tsuruta 1998[21].

In Fig. 1a the dotted and solid curves refer to thermal evolution
of a 1.4M$_\odot$ NS with and without heating, respectively. Since
for these stars the central density $\rho^c < \rho_{tr}^Y$, they
consist predominantly of neutrons and they cool by the slower
`standard' processes. The dot-dashed and dashed curves present
cooling of 1.6M$_\odot$ and 1.8M$_\odot$ hyperon stars. For the
TNI6U EOS adopted, we find that $\rho^c$ = $\rho_{tr}^Y$ for a
1.5M$_\odot$ star.  Therefore, our stars with mass larger than
$\sim$ 1.5M$_\odot$ contain a hyperon core and hence the
predominant cooling mechanism is the nonstandard hyperon direct
Urca process. However, the 1.6M$_\odot$ star (dot-dashed) does not
cool much faster than less massive neutron stars because the
superfluid suppression is very large. The 1.8M$_\odot$ star
(dashed) cools faster because the superfluid gap decreases
significantly for this larger mass (and hence denser)
star\footnote{Note that after the superfluidity sets in the gap
first increases, reaches a peak and then decrease to zero as
density increases.}. See T06a[10] for further details.

Fig. 1b (the right panel) shows thermal evolution of neutron stars
with a pion core. The EOS adopted is `TNI3P Model', which is a
modified version of TNI3 EOS for neutron matter recently
constructed[6]. This EOS is somewhat stiffer than medium. It was
modified by T06b[11], to include pions for densities exceeding
$\rho_{tr}^\pi$, the critical density for transition to pions. The
pion transition density $\rho_{tr}^\pi$ is set to be 2$\rho_0$,
significantly lower than that for hyperons, adopting the results
of recent careful theoretical studies[8]. As the superfluid model
for the pion-condensed phase we adopt the result from the most
recent calculations by Tamagaki and Takatsuka (2006)[8] which
indicates that the gaps for pion condensates are significantly
larger for a significant range of densities above 2$\rho_0$,
although we assume that they decrease at higher densities. Other
microphysical input is the same as in Fig. 1a.

In Fig. 1b the dotted and solid curves refer to thermal evolution
of 1.4M$_\odot$ pion stars with and without heating, respectively.
Since the transition density is low for pions we find that the
central density of a 1.4M$_\odot$ star already exceeds the pion
transition density even for this relatively stiff EOS chosen, and
hence its core already consists of pion condensates. However, for
these stars the gap, and hence T$^{cr}$, is so large that the
superfluid suppression is essentially complete, which means the
curves lie essentially in the same positions as the standard
cooling. The dot-dashed and dashed curves present cooling of
1.6M$_\odot$ and 1.8M$_\odot$ pion stars, respectively. For the
higher central density of these more massive stars the superfluid
gap decreases and hence these stars cool faster. However, for
these stars the gap is still large enough to keep the superfluid
suppression significant. See T06b for further details.

\section{Comparison with Observation}

In Fig. 1 thermal evolution curves are compared with the latest
observational data. We may note that the data suggest the
existence of at least two classes of sources, hotter stars (e.g.,
(F) RX J0822-4300, (G) 1E1207.4-5209, (2) PSR 0656+14, (N)
J1856-4754, and (4) PSR 1055-52), and cooler stars (e.g., (C) PSR
J0205+6449 in 3C58, (1) the Vela pulsar and (I) PSR 1046-58). The
hotter sources are consistent with thermal evolution of less
massive stars such as 1.4M$_\odot$ stars. For PSR 1055-52(4) the
age uncertainty is relatively large, but still it will probably
require at least moderate heating. Source (F) is slightly above
the dotted curves, but for this source the distance is quite
uncertain and if it is closer that will bring the data point down.
Also if the star has magnetic envelopes with light elements such
as Hydrogen cooling will be slower and that will bring the curves
somewhat higher[22]. Note that for this source the best fit to the
spectral data requires a Hydrogen atmosphere[10].

Comparison of cooler star data with pion and hyperon curves
confirms the earlier conclusion[3][21][23] that nonstandard
cooling of more massive stars is required for these cooler data.
In this case we find that significant superfluid suppression is
required, at least for (1) the Vela pulsar detection data. The age
uncertainty should not affect this conclusion, especially for
younger cooler sources such as the pulsars in (C) 3C58 and (1)
Vela, because the slope of the curves in these younger years is
relatively flat. Also the uncertainty for the age estimate from
pulsar spins is small for younger pulsars, at most a factor of 2
or so. Note that for the pulsar in 3C58 the difference between the
age estimates coming from the SNR data and pulsar data is very
small[10].

Recently a possibility for very cold NSs in at least four
supernova remnants (SNRs) was reported by Kaplan et al. 2004[24].
These authors note that if there is a NS in these SNRs the upper
limits to their surface luminosity should be very
low\footnote{these authors pointed out that although some of these
SNRs may contain no compact collapsed objects or the compact
remnants may be black holes, it is quite unlikely that none of
them contains a NS.}. We do not place these upper limits in our
Fig. 1 because their data are given as L$_x$(0.5 - 10 keV), the
X-ray luminosity within the limited window between 0.5 - 2 Kev.
That should be significantly lower than L$_{bol}$, the total
bolometric luminosity over all wavelengths, which should be the
one to be compared with theoretical luminosity in the cooling
curves. For instance, L$_{bol}$ $\sim$ 80 L$_x$(0.5 - 10 keV) for
PSR 0656+14 (the former, $\sim$ 8 x 10$^{32}$ ergs/s, vs the
latter, $\sim$ 10$^{31}$ ergs/s). However, even so, we note that
these upper limits are safely below the standard cooling curves,
and hence a nonstandard fast cooling scenario is required.

\section{Discussion}

Until recently it was thought that at least for binary pulsars
observations offered stringent constraints on the mass of a NS, to
be very close to 1.4M$_\odot$[21]. If this evidence extends to
isolated NSs also, then the EOS should be severely constrained
because it has to be such that the mass of the star whose central
density is very close to the transition density (where the
nonstandard process sets in) should be very close to
1.4M$_\odot$[21]. Very recent observational data, however, suggest
that the mass range should be much broader, $\sim$ 1 --
2M$_\odot$[9][25]. If so, that still should give some useful
constraint on the EOS. For instance, a very soft EOS, such as the
BPS Model[21], should be excluded because for this EOS the maximum
mass is only $\sim$ 1.5M$_\odot$, and hence stars with mass larger
than this cannot be explained. That is why we chose medium to
stiff EOSs for our models. We chose a stiffer EOS for pion stars
because the transition to pions takes place at lower densities.

The qualitative behavior of all nonstandard scenarios is similar
if their transition density is the same[21]. However, here we try
to demonstrate that it is still possible to offer comprehensive
assessment of at least which options are more likely while which
are less likely. First of all, we note that all of the nonstandard
mechanisms without suitable suppression are too fast for all the
detection data. Significant suppression of neutrino emissivity due
to superfluidity is required, to be consistent with cooler stars
such as the Vela pulsar. However, Takatsuka and Tamagaki (1997),
hereafter TT97[26] already showed, through careful microphysical
calculations, that for neutron matter with such high proton
concentration as to permit the nucleon direct Urca, the superfluid
critical temperature T$^{cr}$ should be extremely low, $\sim$
several x 10$^7$ K, not only for neutrons but also for protons.
Here we emphasize that this conclusion does not depend on the
nuclear models adopted for the calculations. On the other hand,
most of the the observed NSs, which are to be compared with
cooling curves, are hotter (the core temperature being typically
$\sim$ 10$^8$ K to several times 10$^8$ K). That means {\it the
core particles are not yet in the superfluid state} in these
observed NSs. Conclusion is that {\it a star cooling with nucleon
direct Urca would be too cold} to be consistent with these
detection data. The same argument applies to kaon cooling
also[27]. Further details are found in [3][10][11][23].

As to the hyperon cooling scinario, we find that that will be a
viable option if recently constructed hyperon superfluid gap
models (which include the Ehime Model adopted here)[7] are valid.
Recently the Gifu-Kyoto nuclear experimental group[28] reported
that the superfluid gap for hyperons would be much smaller. If so,
hyperon cooling also would be in trouble. However, since then
their experiment has not been confirmed by follow-up experiments.
We find that the pion cooling option is still valid.

A few other groups have calculated neutron star cooling. Due to
lack of space here we comment on only the recent major work by
Yakovlev et al 2004, hereafter referred to as Y04[16]. Although
these authors sometimes adopted simplified `toy models' with the
isothermal and other various approximations, their results and
ours generally agree, at least qualitatively, when similar input
is applied.
There are, however, some serious differences. For instance,

(i) In an effort to bring up the standard cooling curves to
explain a hot pulsar PSR 1055, these authors conclude that neutron
superfluidity must be so weak as to be negligible. However, this
conclusion contradicts with the results of serious theoretical
studies of neutron superfluidity, which find that neutron
superfluidity could not be so small for normal neutron matter
where proton concentration is small[6]. On the other hand, we have
shown (see Fig. 1) that this apparent discrepancy disappears even
with models with significant neutron superfluidity when heating is
included. Also, it may be pointed out that the age uncertainty is
rather large for this pulsar since it is older[10]. Therefore, the
unrealistic assumption of negligible neutron superfluidity for
normal neutron matter is not required.

(ii) To explain cooler stars, these authors chose the nucleon
direct Urca process, which they called Durca, as their nonstandard
cooling option. Also, to explain both hot PSR 1055 and cooler Vela
pulsar Y04 require their models to possess large proton
superfluidity and yet negligible neutron superfluidity. However,
TT97[26] already showed that that is impossible for models where
the Durca option works. Specifically, for Durca to operate proton
concentration must be significant. In such a case TT97 showed that
both neutron and proton superfluidity must be very small - it
cannot be that one is very high while the other negligibly small.
In other words the models by Y04 are unphysical. Also, for such
models where Durca can operate it is impossible for proton
superfluidity to be strong enough to offer sufficient suppression
to explain the Vela pulsar. Here we emphasize that their models
are based on studies of normal neutron matter where proton
concentration is small, while that argument breaks down for
special models with high proton concentration which will allow
nucleon Durca to operate.

\section{Summary and Concluding Remarks}

We have shown that the most up-to-date observed temperature data
are consistent with the current thermal evolution theories of
isolated NSs if less massive stars are warmer while more massive
stars are cooler. The comparison of theory with observation,
especially with the low temperature upper limit for the pulsar in
3C58, shows that fast nonstandard cooling is required for cooler
stars. The need for nonstandard cooling is further strengthened by
the recent report by Kaplan et al.[24] for the very low upper
limits to neutron stars possibly present in some of four SNRs.

Among various nonstandard cooling scenarios, both nucleon Durca
and kaon cooling may be excluded. The major reason is that for
nucleon Durca to be operative, high proton concentration is
required, which weakens superfluidity of both protons and
neutrons. Then nucleon Durca will be too fast to be consistent
with e.g., the Vela pulsar data.  Similar argument applies to kaon
cooling. Hyperon cooling may be in trouble if the hyperon
superfluid gap should be so small as reported by recent nuclear
experiments, although this report is yet to be confirmed by
follow-up experiments. On the other hand, pion cooling is still
consistent with both observation and theory. The conclusion is
that {\it the presence of `exotic' particles, possibly pion
condensates, will be required within a very dense star.} If the
need for larger mass stars most recently reported for binary
pulsars is confirmed, the very soft EOSs should be excluded.

The capability of constraining the composition of NS interior
matter purely through observation alone will be limited, due to
often very large uncertainties, mainly for stellar distance and
age. Therefore, it will be very important to {\it exhaust all
theoretical resources.} Theoretical uncertainties are also very
large, especially in the supranuclear density regime. However,
here we emphasize that we should still be able to set {\it
acceptable ranges} of theoretical feasibility, at least to
separate models more-likely from those less-likely.


\begin{theacknowledgments}
We acknowledge with special thanks the contributions by our
collaborators, M.A. Teter, J. Sadino, J. Thiel, T. Takatsuka, R.
Tamagaki, W. Candler, K. Nomoto, H. Umeda, A. Liebmann, and K.
Fukumura. Thanks are due to K. Nomoto, T. Tatsumi and others in
their groups for their hospitality and help during our visits to
Tokyo University and Kyoto University, and the participants of
Kyoto workshops for valuable discussions. Our work for this paper
has been supported in part by NASA grants NAG5-3159, NAG5-12079,
AR3-4004A, and G02-3097X.
\end{theacknowledgments}



\end{document}